\documentclass[letterpaper,twocolumn,10pt]{article}
\usepackage{usenix,epsfig,endnotes,url,mathtools,tabulary,verbatim,graphicx,array}
\usepackage[skip=10pt,font=footnotesize,labelfont=bf]{caption}
\begin{document}

\setcounter{table}{0}
\setcounter{section}{0}

%don't want date printed
\date{}

%make title bold and 14 pt font (Latex default is non-bold, 16 pt)
\title{\Large \bf Humans can discriminate trillions of olfactory stimuli, or more, or fewer}

\author{
{\rm Richard C. Gerkin}\\
Arizona State University\\
School of Life Sciences\\
rgerkin@asu.edu
\and
{\rm Jason B. Castro}\\
Bates College\\
Department of Psychology\\
Program in Neuroscience \\
jcastro@bates.edu
}

\maketitle

% Use the following at camera-ready time to suppress page numbers.
% Comment it out when you first submit the paper for review.
\thispagestyle{empty}

\subsection*{Abstract}
A recent \textit{Science} paper \cite{bushdid_humans_2014} proposed that humans can discriminate between at least a trillion olfactory stimuli. 
Here we show that this claim is the result of a fragile estimation framework capable of producing nearly any result from the reported data, 
including values tens of orders of magnitude larger or smaller than the one originally reported in \cite{bushdid_humans_2014}. 
We conclude that there is no evidence for the original claim.  

\section{Introduction}
\label{sec:intro}
A recent paper \cite{bushdid_humans_2014} proposed that humans can discriminate between at least a trillion olfactory stimuli. 
Using that paper's methods to reanalyze the data it presented, 
we show that this estimate is problematically fragile. 
Specifically, it varies systematically and sensitively 
(over tens of orders of magnitude, in both directions), 
for modest changes in incidental experimental and analysis parameters against which a result ought to be robust.  
Had the experiment enlisted $\sim100$ additional subjects similar to the original ones, 
the same analysis would have concluded that \textit{all possible stimuli} are discriminable 
(i.e. that each of the more than $10^{29}$ olfactory stimuli possible in their framework are mutually discriminable). 
By contrast, if the same experimental data were analyzed using moderately more conservative statistical criteria, 
it would have concluded that there are fewer than 5,000 discriminable olfactory stimuli -- 
no larger than the folk wisdom value that the new estimate purports to replace. 

As a result, data describing the same underlying perceptual abilities admit a wide range of extremely disparate, 
yet unobjectionable alternative conclusions (including both the largest and smallest possible estimates allowed by the analysis framework). 
We conclude that the framework is unsound; 
therefore there may be trillions of discriminable odor stimuli, or more, or fewer, 
but the framework is incapable of settling this issue.  
Here we first demonstrate the framework's fragility, and then explain its origin. 
For most of this paper, we remain agnostic about whether the framework is conceptually sound, 
to highlight the fact that it has strictly methodological problems of a statistical origin that do not depend on the validity of a competing set of assumptions.
In a concluding section, we explore possibilities for improving the estimate. 

\begin{figure}[hbt]
    %\centering
    \includegraphics[width=0.45\textwidth]{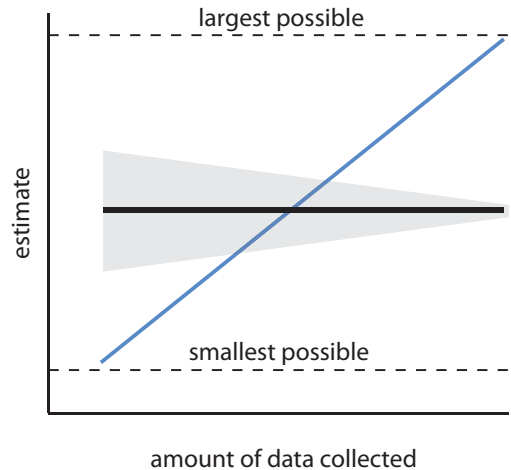}
    \caption{
Consistency of an estimator. 
An estimator is consistent if the resulting estimate asymptotically converges (in expectation) 
as sample size increases (black line). 
Uncertainty in the estimate (gray area) may shrink with sample size, 
but the estimate itself should not systematically change with sample size, and should converge on the truth. 
Estimators without this property are termed \textit{inconsistent} (the blue line is a relevant example), and are considered unreliable, as the resulting estimate can be heavily biased by the sample size.  If the estimate has a minimum and maximum allowed value (see equation \ref{eq:disc}), an especially inconsistent estimator can even produce any estimate within that range.  
}
    \label{fig:statpathology}
\end{figure}

\section{Problems with the estimate}
\label{sec:demonstration}

The main concern is that the estimated number of discriminable stimuli depends steeply, systematically, and non-asymptotically on choices of arbitrary experimental parameters, 
among them the number of subjects enrolled, the number of discrimination tests performed, and the threshold for statistical significance.  
We show below that the order of magnitude claim of `at least one trillion olfactory stimuli' requires that those parameters assume a very narrow set of values. 
Certainly, the precise value of an estimate may change as additional data are collected, 
but the estimate should not change \textit{in expectation}; it should not be possible to make an estimate arbitrarily large (or small), 
simply by collecting more (or less) data. Similarly, the estimate itself should not become arbitrarily small or large with adjustment of a significance criterion. Estimates that scale systematically with such incidental parameter choices are considered statistically \textit{inconsistent} (Figure \ref{fig:statpathology}). It is the inconsistency of the present estimate that produces a tremendously large space of extremely different, yet unobjectionable alternative conclusions that can be reached about the number of discriminable olfactory stimuli. 

\begin{figure}[!hbt]
    \centering
    \includegraphics[width=0.475\textwidth]{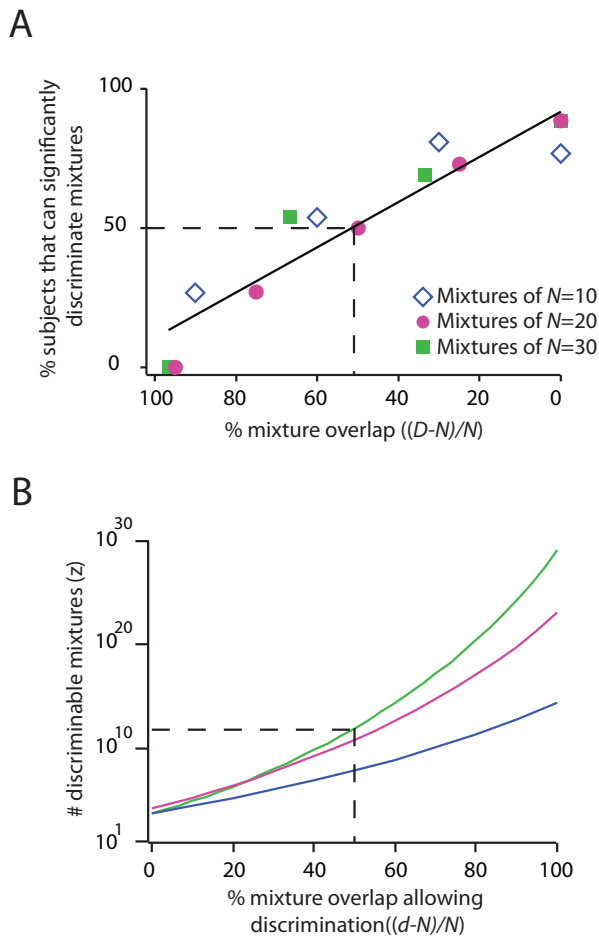}
    \caption{
Reproduction of the main result published in \cite{bushdid_humans_2014}, 
from analysis of raw data made available in supplemental materials of \cite{bushdid_humans_2014}. Compare to Figures 3 and 4 in that publication. 
\textbf{A}: Discriminability vs. Mixture overlap, expressed as a percentage of the mixture size $N$. 
From this analysis, \cite{bushdid_humans_2014} derives $\frac{d-N}{N}\sim51\%$ (vertical dashed line) as the critical value of mixture overlap at which $50\%$ of mixtures achieve `significant discriminability'. 
\textbf{B}: Estimated number of discriminable mixtures $z$ vs. mixture overlap (expressed as a percentage of $N$) allowing discrimination. 
The plot is obtained by regression and interpolation of results like in \textit{A} combined with equation \ref{eq:disc}. 
For a value of $\sim51\%$ as obtained in \textit{A}, 
one obtains the `trillions' figure reported in \cite{bushdid_humans_2014}.}
    \label{fig:reconstruction}
\end{figure}

To illustrate that we can correctly recapitulate the analysis undertaken in \cite{bushdid_humans_2014}, 
Figure \ref{fig:reconstruction} shows our reproduction (using raw supplementary data) of two critical figures from that paper \cite{bushdid_humans_2014}, from which its main conclusion was drawn. 
Figure \ref{fig:colormaps} and Table \ref{table:possible_results} quantify the fragility of this conclusion, 
by generating estimates using the same framework under trivial alternative scenarios in which different numbers of subjects (or mixtures) were used, 
or different choices of statistical threshold ($\alpha$) were used for assessing discriminability. 
See Table \ref{table:definitions} for definitions of parameters used here and in \cite{bushdid_humans_2014}.  
Thus, we produced all values shown here by analyzing the data from \cite{bushdid_humans_2014}, using the methods described therein, and varying only parameters. 
Code for these and all subsequent analyses are available at \url{http://github.com/rgerkin/trillion}. 

\newcolumntype{C}[1]{>{\centering\arraybackslash}m{#1}}
\begin{table}[hbtp]
\caption{Definitions of parameters}
\label{table:definitions}
\begin{tabulary}{0.5\textwidth}{| C{0.6cm} | m{6.35cm} |}
\hline
$z$ & \footnotesize Estimated number of discriminable olfactory stimuli \\
\hline
$C$ & \footnotesize Number of distinct compounds available to make mixtures \\
\hline
$N$ & \footnotesize Number of distinct compounds in a mixture \\
\hline
$O$ & \footnotesize Number of distinct compounds shared by a mixture pair \\
\hline
$D$ & \footnotesize Number of distinct compounds in one mixture of a pair that are not shared by the other.  ($D = N-O$) \\
\hline
\emph{class} & \footnotesize All mixture pairs with the same value of $N$ and $D$. \\
\hline
$d$ & \footnotesize The value of $D$ for which mixture pairs of a given $N$ are more likely than not to be discriminable at a rate significantly above chance. \\
\hline
\end{tabulary}
\end{table}

\begin{figure*}
    \centering
    \includegraphics[width=1.0\textwidth]{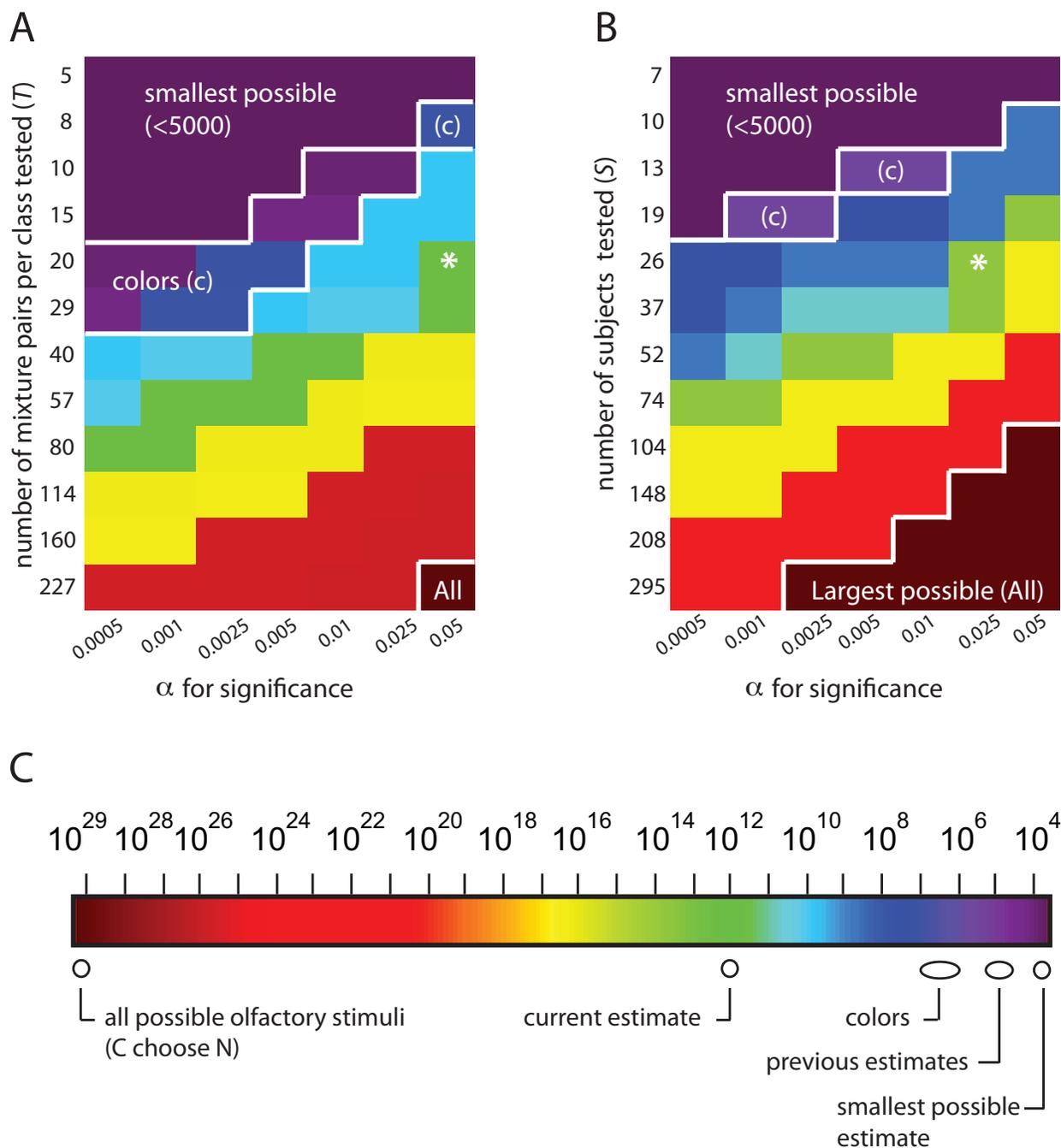}
    \caption{
The estimation framework supports nearly any alternative conclusion, including the smallest and largest estimates possible under the framework. 
\textbf{A}: Heat map showing alternative conclusions reached for different choices of $T$, the number of mixture pairs per class to test, 
and application of alternative significance threshold $\alpha$ for discriminability, with the data from \cite{bushdid_humans_2014}. 
Asterisks (*) show the parameter regime ($T=20$ mixtures, $\alpha=0.05$) used in \cite{bushdid_humans_2014}. 
Other values on each axis are chosen in a geometric progression around those parameters.  
The contour in the lower right labeled `All' demarcates a regime in which one will conclude that the largest possible number of of mixture stimuli (i.e. all $z(d=0)={128 \choose 30} > 10^{29}$ of them) are discriminable (see equation \ref{eq:disc}). 
The contour in the upper left labeled `smallest possible' demarcates a regime in which one will conclude that 
the smallest possible number of stimuli are discriminable, i.e. only $z(d=N=30)<5000$ of them.
The contour labeled `colors' demarcates a regime in which 
one concludes that the number of discriminable olfactory stimuli is the same order of magnitude as the number of discriminable colors. 
\textbf{B}: Heat map similar to left, only with number of subjects on the vertical axis.  
A choice of $\alpha=0.025$ is necessary to obtain the estimate that \cite{bushdid_humans_2014} reports for this analysis.
\textbf{C}: Colorscale for \textit{A} and \textit{B}, with reference landmarks.}
    \label{fig:colormaps}
\end{figure*}

\renewcommand{\arraystretch}{1.2}
\newcolumntype{C}[1]{>{\centering\arraybackslash}m{#1}}
\begin{table}[!hbt]
\centering
\begin{flushleft}
A
\end{flushleft}
\begin{tabulary}{0.02\textwidth}{|C{2.3cm}|C{2cm}|C{1.6cm}|}
\hline
\# Discriminable stimuli ($z$) & Significance threshold ($\alpha$) & \# Tests per class ($T$) \\
\hline
$2.02*10^{12}$ & 0.05* & 20*\\
\hline
$4.56*10^{3}$\textdagger & 0.05* & 5\\
\hline
$1.54*10^{29}$\textdaggerdbl & 0.05* & 185\\
\hline
$8.94*10^{3}$ & 0.001 & 20*\\
\hline
$1.79*10^{4}$ & 0.01 & 15\\
\hline 
\end{tabulary}
\linebreak
\begin{flushleft}
B
\end{flushleft}
\begin{tabulary}{0.02\textwidth}{|C{2.3cm}|C{2cm}|C{1.6cm}|}
\hline
\# Discriminable stimuli ($z$) & Significance threshold ($\alpha$) & \# Subjects ($S$) \\
\hline
$3.81*10^{13}$ & 0.025* & 26*\\
\hline
$4.56*10^{3}$\textdagger & 0.025* & 7\\
\hline
$1.54*10^{29}$\textdaggerdbl & 0.025* & 135\\
\hline
$3.47*10^{7}$ & 0.001 & 26*\\
\hline
$2.98*10^{5}$ & 0.01 & 15\\
\hline 
\end{tabulary}
\caption{Estimates of $z$, the number of discriminable olfactory stimuli, for different possible parameters values, for the $C=128$, $N=30$ case used in \cite{bushdid_humans_2014}. This recapitulates selected points from Figure \ref{fig:colormaps}. 
* indicates that the parameter value was used in \cite{bushdid_humans_2014}. 
We assume here that new subjects perform similarly to the original subjects. 
Note that $4.56*10^3$ (\textdagger) and $1.54*10^{29}$ (\textdaggerdbl) are the smallest and largest possible values allowed by the framework from \cite{bushdid_humans_2014}.}

\label{table:possible_results}
\end{table}

In \cite{bushdid_humans_2014}'s experimental framework, there are three sets of experiments, varying in the number of distinct molecular components $N$ per mixture tested.  
We consider the $N=30$ case (without loss of generality) for which there are $\sim 10^{29}$ possible olfactory stimuli, 
and for which the smallest possible number of discriminable stimuli is $\sim 4500$ (see equation \ref{eq:disc}, in section \ref{sec:explanation} below).  
Figure \ref{fig:colormaps} and Table \ref{table:possible_results} thus demonstrate that \textit{1)} there is a regime of reasonable parameter choices for which one concludes that all possible olfactory stimuli (i.e. all $\sim 10^{29}$ of them) are discriminable; 
and \textit{2)} there is another regime of reasonable parameter choices for which one concludes that the smallest possible number of stimuli (i.e. only $\sim 4500$) are discriminable.  
The only assumption required to obtain these estimates is that performance in new subjects is similar to performance in the original subjects.  

The fragility of the conclusion results from the claim in \cite{bushdid_humans_2014} that a modest (if very interesting) correlation -- between the discriminability of a pair of mixtures and the overlap (fraction of shared components) of those mixtures --
is evidence that a \emph{particular degree} of mixture overlap defines a boundary that partitions the discriminable from the indiscriminable in a very high-dimensional space.  
Below, we explore the consequences of this decision, and its implications for calculating the number of discriminable olfactory stimuli. 

\section{Explanation of the problems with the estimate}
\label{sec:explanation}

\subsection{Recap of the basic framework}
\label{sec:recap}

\begin{figure}[!hbt]
    \centering
    \includegraphics[width=0.475\textwidth]{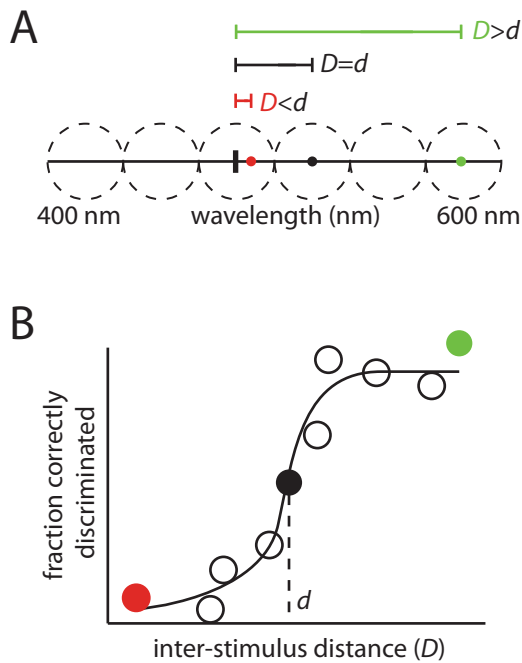}
    \caption{
`Sphere packing' to estimate the number of discriminable colors: the motivation behind the framework in \cite{bushdid_humans_2014}. 
\textbf{A}: Hypothetical example showing a range of visible wavelengths. 
Relative to a reference stimulus (thick vertical tick mark), extremely distant stimuli (green circle) in this space are easy to discriminate, 
whereas extremely close stimuli (red circle) may be impossible to discriminate, 
as they are beyond the resolution of color vision. 
At some critical inter-stimulus distance, $d$, 
stimuli will be `just discriminable' (black circle). 
A typical stimulus pair on the space, separated by distance $D$, 
will tend to be discriminable if $D>d$, and indiscriminable if $D<d$. 
\textbf{B}: This partitioning into discriminable and indiscriminable sets is captured in the sigmoidal shape 
of the psychometric curve plotting discriminability vs. distance. 
Knowing that an interval of length $d$ on the space will tend to span `just discriminable' stimuli, 
one can calculate how many such intervals, $z$, can be `packed' onto the space
to estimate the number of discriminable colors.}
    \label{fig:spherepacking}
\end{figure}

The framework's logic is built on an analogy to color vision, 
where estimating the number of discriminable colors requires knowing only two numbers: 
the size of the stimulus space (that is, the range of visible wavelengths), 
and the minimally discriminable distance between a typical pair of stimuli (Fig. \ref{fig:spherepacking}). 
Dividing the first number by the second amounts to asking how many discriminable intervals can be `packed' into the stimulus space, 
with that number providing an estimate of the number of discriminable color stimuli. 

Because olfactory stimuli do not have obvious physical dimensions analogous to wavelength, 
olfaction is not amenable to an identical calculation.  
Instead, \cite{bushdid_humans_2014} established a theoretical framework that yielded a similar calculation based upon the same underlying idea.  
\cite{bushdid_humans_2014} proposed to divide the size of a investigator-determined olfactory stimulus space by a data-determined variable representing resolution in this space. 
Instead of being continuous, one dimensional, and defined by some intrinsic stimulus variable like wavelength, 
the olfactory stimulus space was defined to be the discrete, high-dimensional space spanned by all mixtures containing $N=30$ different components (molecules) that could be assembled from a library of $C=128$ molecules; 
\cite{bushdid_humans_2014} also considers the $N=10$ and $N=20$ cases, which we ignore in this section with no loss of generality. 
This space of possible mixture stimuli is astronomically large ${C \choose N}$, 
owing to the proverbial `combinatorial explosion', 
and each point in the space corresponds to a specific multi-component mixture. 

One definition of distance between stimuli in this space is the number of components $D$ by which the stimuli differ. 
For example, nearest neighbors would be stimuli sharing all components but one ($D=1$), 
and the most distant points in this space would be stimuli differing in all components ($D=N$). 

\cite{bushdid_humans_2014} showed that discriminability of a stimulus pair tends to increase with the distance $D$ between the stimuli in that pair (Figure \ref{fig:reconstruction}A), 
and then argued for the existence of a special distance $d$ corresponding to the $D$ at which stimuli are `just discriminable'.  
In other words, for $D>d$ stimuli should more often than not be considered discriminable 
and for $D<d$ they should more often than not be considered indiscriminable.  
By calculating $d$, one could in turn readily calculate the number of stimuli within a distance $D \leq d$ of a typical point in the stimulus space using the provided formulas. 
Geometrically, the set of stimuli with distance $D \leq d$ from a reference stimulus corresponds to a filled `ball' of stimuli indiscriminable from the reference stimulus at its center.  
Conversely, the reference stimulus should be discriminable from stimuli outside the ball.  
We could thus count the number $z$ of non-overlapping balls that can be packed into the stimulus space, 
as the proposed in \cite{bushdid_humans_2014}, by analogy to the example for color vision: 

\begin{equation}
\label{eq:disc}
z(d) = \frac{\binom{C}{N}}{ball(d/2)}  
\end{equation}

where `ball' is defined as: 

\begin{equation}
\label{eq:ball}
ball(r) = \sum_{x=0}^{r} \binom{N}{x}\binom{C-N}{x}
\end{equation} 

Equation \ref{eq:disc} produces the final estimate $z$ of the number of discriminable stimuli. 
$C$ and $N$ are fixed by experimenter choices, 
and $d$ -- the resolution-like term -- is the only quantity derived from data that is related to measured psychophysical performance. 
Note that for $C=128$, $N=30$, as used in \cite{bushdid_humans_2014}, the \textit{largest} and \textit{smallest} possible values this equation can produce are $\sim 1.5 * 10^{29}$ (for $d=0$) and $\sim 4500$ (for $d=N$), respectively.  
Assuming this framework is conceptually unproblematic (but see \cite{meister_can_2014}), 
the only question becomes: How do we derive $d$ from the data? 

\begin{figure}[!hbt]
    \centering
    \includegraphics[width=0.475\textwidth]{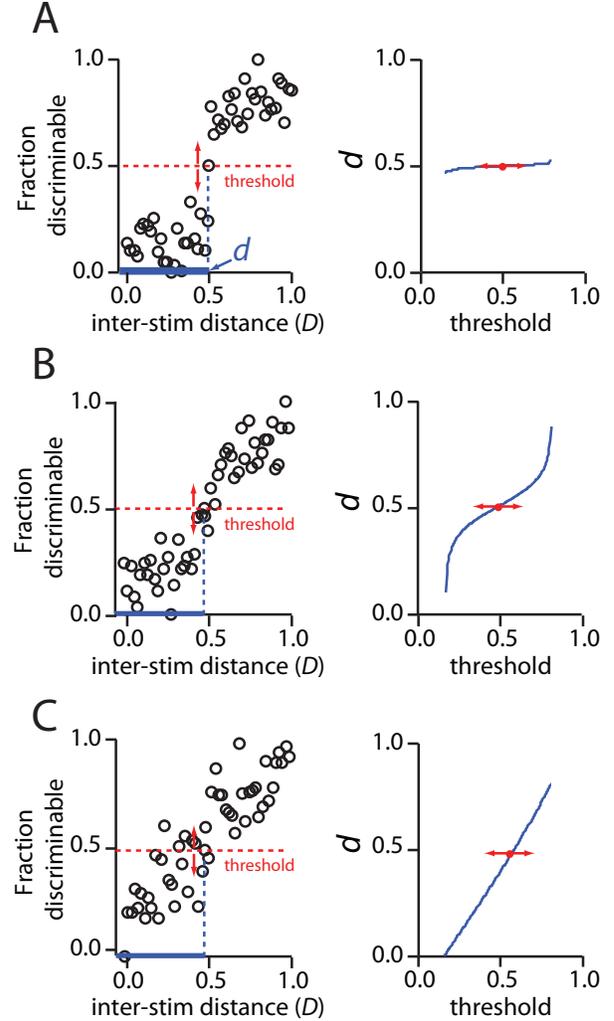}
    \caption{
Behavior of psychometric curves for hypothetical data describing discriminability vs. inter-stimulus distance. 
\textbf{A}: \textit{Left}, A sharply sigmoidal relationship in which 
discriminability changes dramatically and categorically at a critical inter-stimulus distance, $d$. In all panels, $d$ is the value of the inter-stimulus distance $D$ at which a threshold fraction of stimulus pairs are discriminable.  In the left panels, this threshold is set at $0.5$.
\textit{Right}, The resulting value of $d$ is nearly invariant to the choice of threshold . 
\textbf{B}: Same as above, only for a less sharply sigmoidal data set. 
There is still a narrow regime in which $d$ is largely invariant to choice of threshold. 
\textbf{C}: Same as above, only for a weakly sigmoidal data set. 
Here, there is no principled means for choosing the $d$ that is characteristic of discriminability relationships for stimuli. 
The data in \textit{C} do not support an interpretation in which 
there is defensible characteristic `length scale' for inter-stimulus distances.}
    \label{fig:sigmoids}
\end{figure}

\subsection{Derivation of the critical parameter $d$} 
\label{sec:derivation}

\subsubsection{Thresholding the fraction discriminated} 

A classic psychometric curve (Figure \ref{fig:spherepacking}B), showing discriminability as a function of inter-stimulus distance $D$, admits a few plausible ways to derive $d$. 
The simplest is to simply use a discriminability threshold, such that $d$ corresponds to the distance $D$ at which the `fraction correct' reaches a certain value.  
In \cite{bushdid_humans_2014}'s three-alternative forced-choice experiments, chance responding would produce a fraction correct of $\frac{1}{3}$, 
so the appropriate threshold would be somewhere between $\frac{1}{3}$ and 1.  
This threshold choice would be arbitrary -- we might say that a fraction correct of $\frac{1}{2}$ reflects discriminable, 
or alternatively we might choose $\frac{2}{3}$ or any other value between $\frac{1}{3}$ and 1.  

\begin{figure}[!hbt]
    \centering
    \includegraphics[width=0.475\textwidth]{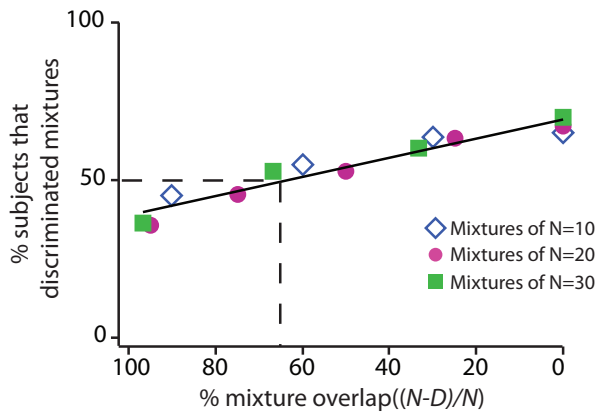}
    \caption{
Discriminability vs mixture overlap.  Analogous to Figure \ref{fig:reconstruction}, except plotting \textit{fraction discriminated} directly (as in Figure \ref{fig:sigmoids}), instead of \textit{fraction significantly discriminable}. The threshold ($50\%$) and the procedure for computing mixture overlap at that threshold are as in Figure \ref{fig:reconstruction}A. Derived from data in \cite{bushdid_humans_2014} as for Figure 2.}
    \label{fig:hardthresh}
\end{figure}

\begin{figure}[!hbtp]
    \centering
    \includegraphics[width=0.475\textwidth]{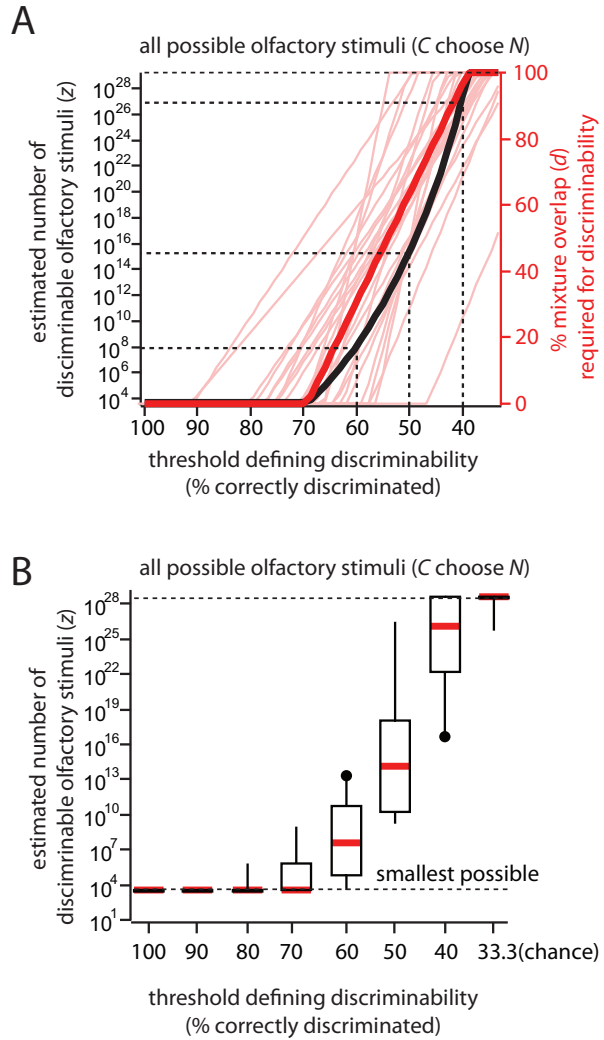}
    \caption{
Relationship between the estimated number of discriminable stimuli $z$ and the choice of threshold defining discriminability. 
\textbf{A}) The thick red line shows the critical distance $d$ that would result from the data in \cite{bushdid_humans_2014} 
for a range of `fraction discriminated' thresholds between $100\%$ (perfect discrimination), 
and $33.3\%$ (chance discrimination). 
The curve was obtained by regression on plots like that in Figure \ref{fig:hardthresh}, by analogy to Figure \ref{fig:reconstruction} and \cite{bushdid_humans_2014}. 
Note that $d$ exhibits a nearly constant-slope relationship with threshold, 
meaning the data are not defined by a characteristic length scale, much like in Figure \ref{fig:sigmoids}C.
The thick black curve shows the relationship between $z$ and the chosen threshold. 
This relationship was obtained directly from $d$, using equation \ref{eq:disc}, as in \cite{bushdid_humans_2014}. 
The thin red lines correspond to the same calculation for $d$ but using data for only a single subject (one per line), 
showing similar sensitivity to the choice of threshold. 
The absence of a robust d for any individual subject argues that the group data are not simply explained by averaging across a population with well-defined, but diverse values of $d$. 
Note that very modest and reasonable alternative choices for the threshold result in extremely disparate estimates. 
The vertical axis is bounded by the smallest and largest possible number of discriminable stimuli allowed by the framework.  
The dashed lines are a visual guide to specific (threshold, $z$) pairs.
\textbf{B}) Box and whisker plots showing the median and inter-quartile range for $z$ when restricting the analysis to individual subjects. 
Note that the worst performing subjects under one threshold can discriminate many more stimuli than the best performing subjects under a slightly more liberal threshold (compare best subject using a $60\%$ threshold vs. worst subject using a $40\%$ threshold).}
    \label{fig:z_d_threshold}
\end{figure}

If the psychometric curve is sufficiently steep near some value of $D$ (Figure \ref{fig:sigmoids}A represents an ideal case) then the derived $d$ will vary minimally over a wide range of choices for the threshold.  
In this scenario, we might be confident that the $d$ we derive is a truly meaningful measure of resolution -- it would be robust.  
If not (Figure \ref{fig:sigmoids}C), it will be very fragile.  
To test whether this scenario applies to the data from \cite{bushdid_humans_2014}, we plotted the fraction discriminated vs percent mixture overlap for that data (Figure \ref{fig:hardthresh}), and then  
varied the threshold from $\frac{1}{3}$ to 1, 
using regression and interpolation to obtain the distance $d$ corresponding to this threshold (Figure \ref{fig:z_d_threshold}, thick red line), 
by analogy to the framework in \cite{bushdid_humans_2014}.  

We subsequently used those values to compute the corresponding number of discriminable stimuli $z$ (Figure \ref{fig:z_d_threshold}, black curve).  
These results show that neither the estimates of $d$ nor by extension that of $z$ are robust across this range of thresholds, 
so it is impossible to report with any confidence the number of discriminable stimuli using this approach. 
Intuitively, this `choose your favorite threshold' strategy is problematic, 
as it effectively amounts to picking a target number between $\sim 10^{3}$ and $\sim 10^{29}$. 
Below, we show that the actual framework used in \cite{bushdid_humans_2014} is nominally employed to make a more principled choice of threshold;
however it merely cloaks the arbitrariness of the threshold choice, but does not eliminate it. 

\subsubsection{Thresholding the fraction \emph{significantly} discriminable} 
A variation on the above approach, and the one used in \cite{bushdid_humans_2014}, 
is to apply a threshold not to the fraction \textit{discriminated}, 
but to the fraction \textit{significantly discriminable}.  
In other words, determine for which subjects (or alternatively, for which classes of mixtures) the fraction discriminated is significantly greater than $\frac{1}{3}$, 
i.e. for which subjects the null hypothesis of chance discrimination can be rejected.  
To facilitate visualization of this step, \cite{bushdid_humans_2014} re-plotted the summary data (fraction correctly discriminated) as fraction significantly discriminable (Figure \ref{fig:reconstruction}A).  
This view of the data again provides a linear relationship between distance $D$ and, in this case, fraction significantly discriminable, 
which holds across all the values of $N$ tested.  
The relationship is now steeper than it was for fraction discriminable (compare Figures \ref{fig:reconstruction} and \ref{fig:hardthresh}) 
because the extra hypothesis-testing step acts as a strong non-linear threshold that exaggerates otherwise small differences in the data.  
Again a choice of threshold choice is required; \cite{bushdid_humans_2014} chose a threshold of $50\%$ significantly discriminable, 
and computed $d$ using linear regression and interpolation as above.  

\begin{figure*}[!hbt]
    \centering
    \includegraphics[width=1.0\textwidth]{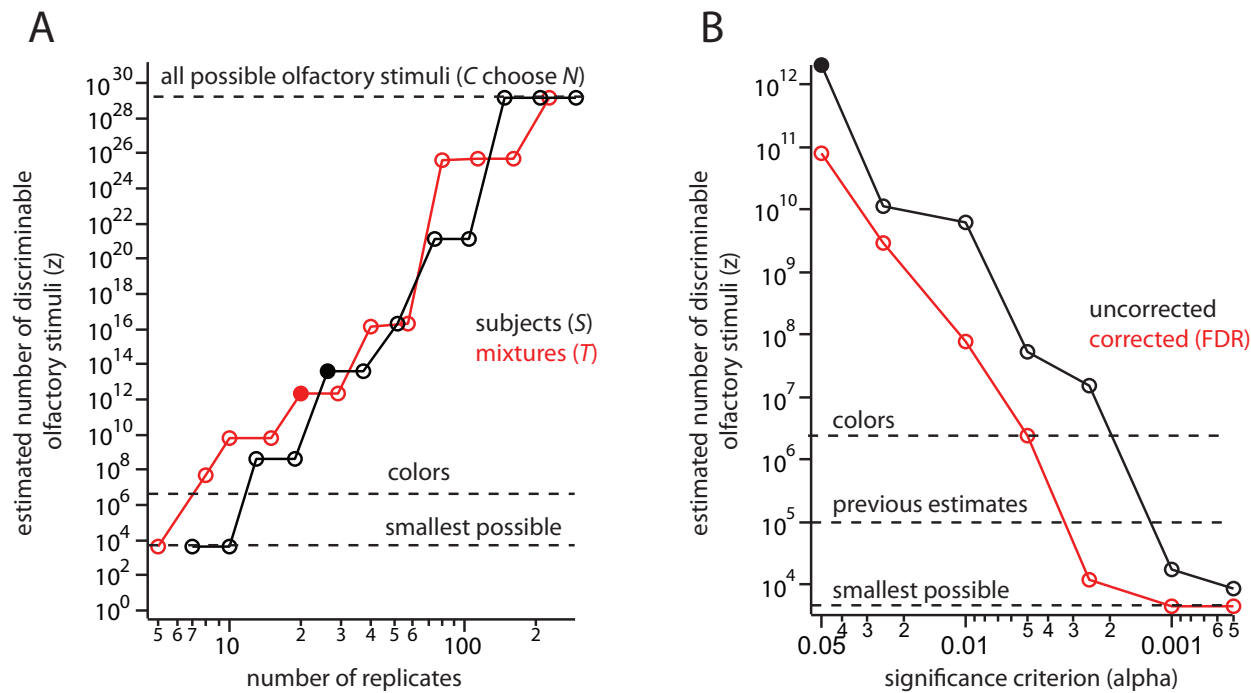}
    \caption{
Steep, systematic, and non-asymptotic dependence of the estimate on sample size ($S$ or $T$) and threshold $\alpha$ for statistical significance. 
\textbf{A}) Dependence of the estimate (for mixtures of $N=30$) on sample size. 
Black shows dependence on the number of subjects $S$ enrolled in the study, 
Red shows dependence on the number of mixtures $T$ tested per mixture class. 
Once the number of mixtures or subjects tested is $\sim 150$ (by no means an unusually large sample size), 
the conclusion that all possible ${C \choose N}$ mixtures are discriminable is guaranteed, in contradiction with experimental results. 
\textbf{B}) Dependence of the estimate on the significance threshold $\alpha$ with (red) and without (black) a correction for multiple comparisons. 
\cite{bushdid_humans_2014} did not correct for multiple comparisons.}
    \label{fig:alpha_dependency}
\end{figure*}

Because the linear relationship between the distance $D$ and the fraction significantly discriminable (Figure \ref{fig:reconstruction}) is steeper than it is for $D$ and the fraction discriminated (Figure \ref{fig:hardthresh}), 
the former would appear to be less fragile than the latter.  
Indeed, varying the threshold (i.e. $50\%$) itself (not shown), 
will have much less effect on the computed $d$ (and consequently on $z$) for the former than for the latter.  
However, by introducing a hypothesis-testing step, the $d$ derived from Figure \ref{fig:reconstruction} now varies systematically with the number of subjects enrolled in the study (and the number of mixtures tested), 
and with the choice of significance criterion $\alpha$.  
This is because each data point used to compute $d$ becomes the binary result of a hypothesis test, 
each of which depends critically on sample size and test specificity.  
Because $d$ is then fed into an expression (equation \ref{eq:disc}) that explodes geometrically, 
the result is a recipe for producing any of a range of estimates for $z$ that one might choose. 
If one enlists more subjects or slackens the significance criterion, 
a very large (even the largest possible) number will be obtained. 
If one enlists fewer subjects or makes the significance criterion more strict, 
a very small (even the smallest possible) number will be obtained. 
Figure $\ref{fig:alpha_dependency}$ shows the explicit dependence of the estimate on these quantities. 
Naturally, these can be varied in tandem too, with even more dramatic consequences, 
as described above (Figure \ref{fig:colormaps} and Table \ref{table:possible_results}). 

A hypothesis test is meant to assess the strength of evidence for or against a hypothesis (often against a null hypothesis), not to make a point estimate.  
However, it may not be uncommon for researchers to use hypothesis testing in the manner done in \cite{bushdid_humans_2014} -- 
to count the number or fraction of data points exhibiting a certain property.  
In many cases this may amount to a venial statistical sin with (hopefully) benign consequences.  
But that is unfortunately not the case here, due in part to the extremely steep dependence of $z$ on $d$ guaranteed by equation \ref{eq:disc}. 

If one claims an estimate to be meaningful, 
it is fair to ask how vigorously would one have to defend a specific choice of arbitrary experimental parameters to defend a particular order-of-magnitude range around that estimate.
Unfortunately, the systematic sensitivities exhibited here severely undermine the plausibility and relevance of the estimate reported in \cite{bushdid_humans_2014}. 
Due to these sensitivities, one could pick almost any number of discriminable stimuli in advance, 
and affirm this number using these or similar data. 
\cite{bushdid_humans_2014} simply exchanged the arbitrariness of a `fraction discriminated' threshold with the arbitrariness of the sample size and $\alpha$. 
Ultimately, the absence of a robust $d$ to characterize the data is an insurmountable obstacle for the framework.   

\section{Building the the stimulus space}
\label{correlations}
\subsection{The structure of the stimulus space}
One might ask: what is the right way to calculate $d$ in order to obtain a robust estimate of the number of discriminable stimuli? 
Before heading down this road and devising alternative statistical approaches, 
it is worth first clearly articulating the assumptions of a framework in which a single variable plays such a special role. 
Under what conditions is it sensible to expect that plugging a single data-derived number into equation \ref{eq:disc} 
will produce a meaningful lower bound of the number of discriminable olfactory stimuli?

To gain some intuition into this, 
we can ask the analogous question in the simplified visual system example (Figure \ref{fig:spherepacking}) 
that was used as the principal motivation for the procedure. 
The `sphere packing' calculation in this case naturally involves measuring the resolution of perception in terms of the stimulus, 
but its validity is not a consequence of this measurement alone. 
Rather, the procedure in Figure \ref{fig:spherepacking} is sensible  because the thing we are calling an independent stimulus dimension (wavelength) is respected as such by perception: 
we encounter monotonically changing, non-redundant percepts as we move from one extreme of the stimulus space to the other. 
If we didn't -- say, if the same percept `blue' were experienced for several non-overlapping disjoint intervals -- 
the sphere packing formulation would fall apart. 
We might observe that on average discriminability improves with distance,
but this would not be evidence of a characteristic length scale 
that partitions stimulus pairs into discriminable vs. indiscriminable sets.

Thus the sphere-packing framework is valid only if the underlying geometry of \emph{stimulus space} 
(that the investigator has designed) aligns with the geometry of \emph{perceptual space} (as implemented in neural circuitry). 
Formally, the map from stimulus space to perceptual space needs to be homeomorphic, or nearly so. 
See \cite{meister_can_2014} for further insight on this issue.  

\subsection{Redundancy in the stimulus space}
Instead of providing evidence for this homeomorphism, \cite{bushdid_humans_2014} assumed for the purposes of calculation that each component of the molecular library 
(of size $C=128$ in \cite{bushdid_humans_2014}) 
spanned an informative additional dimension for perception to explore: 
each molecule in the library is treated as an olfactory primary that is independent of all the others. 
This is the assumption, codified in the numerator of equation \ref{eq:disc}, 
that allows for a massive space of potential discriminable stimuli. 
Indeed, the guaranteed runaway growth of the numerator as molecules are added to the $C$-sized library was offered in  \cite{bushdid_humans_2014} as an argument for why the reported `trillion' figure is a lower bound -- after all, $C$ could always be higher. 

\begin{figure}[!hbt]
    \centering
    \includegraphics[width=0.475\textwidth]{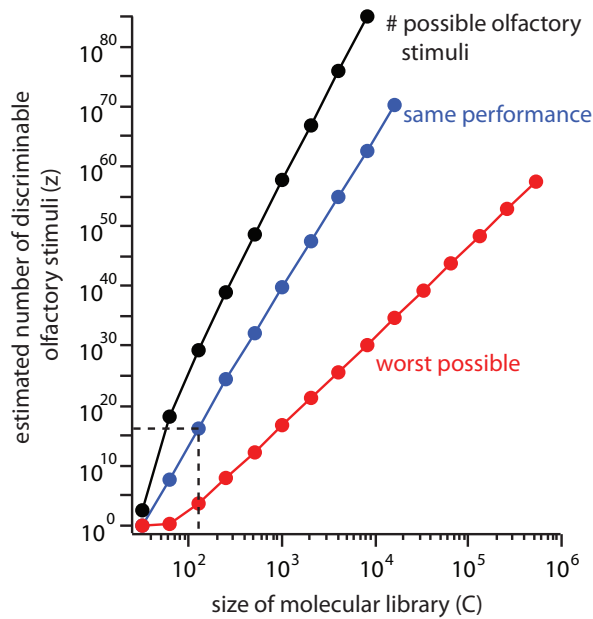}
    \caption{
Explosive growth of the estimate $z$ on the size ($C$) of the molecular library.  The number of possible stimuli $z$ that can be assembled by choosing $N=30$ distinct molecules from a library of size $C$ increases geometrically with $C$ (black line).  If a library of a different size had been used, and similar subject performance resulted, the estimated number of discriminable stimuli $z$ would grow along a similar trajectory (blue line).  Even if performance deteriorated as $C$ increased, the estimate could never fall below the red line, which represents worst-case performance ($d=N$). This results from the combinatorial explosion inherent in equation \ref{eq:disc}.}
    \label{fig:z_vs_C}
\end{figure}

It is worthwhile to quantify the behavior of the estimate as $C$ changes. 
First, the estimate depends geometrically on $C$, 
with a power law exponent of $\sim 30$ (Fig. \ref{fig:z_vs_C}, blue line). 
In other words, if the chemical library were doubled, 
the estimate $z$ would increase by a factor of $2^{30}$ under constant performance. 
If the component library were increased to the size of a standard flavor and fragrance catalog ($\sim2000$ chemicals), 
the estimate would increase to $z\sim10^{41}$, 
implying a unique olfactory percept for each carbon atom on earth. 

Subjects' performance could become worse when mixtures are drawn from this larger, more complete library, 
and we acknowledge that we cannot know in advance what the newly calculated resolution $d$ would be on the new stimulus space.  
In other words, as the numerator of equation \ref{eq:disc} increased, 
its denominator (given by equation \ref{eq:ball}) might conveniently grow proportionally.  
Let us therefore assume that with a library of sufficient size, 
so many mixtures become indiscriminable that the resolution becomes as poor as the framework allows, with $d=N$. 
Even in this edge case, if only mixtures differing in all components were `just discriminable', 
we would still calculate $10^{21}$ discriminable stimuli. 
If $C$ is increased to $10^6$, 
the smallest possible number of discriminable percepts 
(under the assumption of worst measurable performance, as above) 
is $10^{61}$, or ten million trillion unique olfactory percepts for every carbon atom on earth (Fig. \ref{fig:z_vs_C}, red line). 
One may object that the inflation of $C$ here is an unfair critique, 
as the perceptual redundancy of molecules must at some point provide an important constraint on the size of the artificially constructed stimulus space. 
Indeed, it has been reported that as few as thirty components are required to imbue most mixtures with a common smell, 
even when there is no component overlap between the mixtures \cite{weiss_perceptual_2012}.
But this is the essence of the problem with equation \ref{eq:disc}: 
where does that point lie, 
and why wasn't the constraint important to consider for the original $C=128$ molecular library?

\section{Avenues for improving the estimate}

If one is seeking a conservative estimate of the number of discriminable stimuli in a perceptual space whose organization and intrinsic dimensionality are poorly understood, it is arguably more appropriate to use a model that accounts for the data with the smallest number of dimensions. The massive estimates possible in the framework are an immediate consequence of a definition of dimensionality driven by experimenter designation, not data. 

We therefore propose an alternative framework: 
use experimental data to create a working map of the perceptual space, 
and then apply the sphere-packing framework to that map, 
rather than to a map of the stimulus space.  
In cognitive science, psychometrics, and marketing, subject responses to stimuli are used 
to create maps of the underlying perceptual (or conceptual) representations of those stimuli.  
These maps are characterized by the attribute that pairs of items which are considered intuitively to be perceptually near 
(rated similar or difficult to discriminate) are nearer to one another on the map than pairs of items 
which are perceptually more distant (rated dissimilar or easy to discriminate).  
There are many algorithms for generating such maps, 
many of which have been used before in olfaction, 
including variants of PCA \cite{khan_predicting_2007,koulakov_in_search_2011,zarzo_identification_2006}, 
non-negative matrix factorization (NMF, \cite{castro_categorical_2013}), 
and multi-dimensional scaling \cite{mamlouk_quantifying_2003}.  
While there are open questions in the generation of these maps 
(e.g. how many dimensions should they have?), 
they all have the virtue that their accuracy can be checked 
(e.g. by examining the correlation between subjects' indications of item pair dissimilarity and the distance between that pair on the map), 
and thus the maps can be improved. 
Developing these maps may also have the collateral benefit of revealing stimulus dimensions intrinsic to olfaction (if any), 
which could constrain the experimental choice of a resolution to measure.

Unfortunately, it is difficult if not impossible to create these maps from the data discussed here, 
because each mixture of a tested pair is used only once in \cite{bushdid_humans_2014}, 
in that pair alone, and never in any other pairs.  
Thus, there are no serial comparisons of the same mixture that could be used to anchor a stimulus on the map 
relative to anything other than that one stimulus against which it was directly compared experimentally.  
Thus, there is no way to compute distances between stimuli that do not appear together in a tested pair.  
In future experiments such serial repetition of already-tested mixtures would be required 
to build up a data set to which the proposed method could be applied.  
\section{Appendix} 

Here, we provide a more detailed statistical argument describing the framework's extreme sensitivity to incidental parameters. 
The crux of the statistical issue is this: 
the framework could only be valid if $d$, the estimated difference limen used in the calculation step, 
is a measure of olfactory resolution that converges to the true value of this quantity as more data is collected, 
i.e. if it is \emph{consistent}. 

`Significantly discriminable' is a moving target dependent on sample size, 
choice of significance criterion, and correction for multiple comparisons. And
 $d$ is the only data-dependent value used in subsequent calculations (equation \ref{eq:disc}), 
Together, this guarantees that the estimate of $z$ in the \cite{bushdid_humans_2014} is a moving target as well, 
dependent on these same parameters.
$d$ is generated by testing a number of null hypotheses, 
and is closely related to the fraction of these which are rejected.  
But the probability of and criteria for rejection of these null hypotheses depends critically on sample size and $\alpha$, the values
that we explored in Figure \ref{fig:colormaps} and Table \ref{table:possible_results}. 
Certainly, we would agree that there is nothing objectionable about the specific parameters chosen in \cite{bushdid_humans_2014}. 
However, there is nothing objectionable about many other values for those parameters either. 

In effect, calculating $d$ is somewhat like judging whether a coin meets a cutoff for being fair based on a series of tosses. 
It matters very much how many tosses one makes, 
and how much deviation from chance one is willing to tolerate before calling a coin unfair. 
If you have no particular reason to believe a coin is unfair, 
you might be disinclined to call it unfair if you observe $\frac{6}{10}$ ($60\%$) heads, 
but probably not if you observed $\frac{600}{1000}$ heads 
(also $60\%$). 
However, if you own a casino, you might call 5100 heads in 10000 ($51\%$) evidence of an unfair coin. 
Whether the coin is fair is not something we directly measure, 
but rather we have more or less evidence for various degrees of fairness.  

A similar situation applies in \cite{bushdid_humans_2014}'s analysis by considering its formal definition of $d$ 
(a definition we verified by reconstructing the critical figures from \cite{bushdid_humans_2014} in Figure \ref{fig:reconstruction}. 
$d$ is defined as that inter-stimulus distance $D$ for which $50\%$ of subjects can significantly discriminate a mixture class. 
By a mixture `class' we denote the set of mixture pairs for which each mixture has the same number of total components ($N$) 
and each pair has the same number of distinct, non-overlapping components $D$ ($D=N-O$, see Table \ref{table:definitions}). 
For example, the mixture pair ($ABC$, $ABD$) would be a member of the class with $N=3$ and $D=1$ distinct components. 
We focus here on calculations pertaining to the number of tests $T$ per class, 
but the same argument is readily translated over to the number of subjects $S$.

To assess significant discriminability from chance, 
\cite{bushdid_humans_2014} used a two-tailed binomial test. 
Thus if a p-value is smaller than $\frac{\alpha}{2}$ then the subject is considered able to significantly discriminate from pairs in the mixture class.  
The p-value is given by 1 minus the cumulative binomial distribution function for $n=T$ trials, $k$ successes, 
and a probability of success equal to $\frac{1}{3}$, 
with $k$ corresponding to the number of subjects discriminating correctly, and $\frac{1}{3}$ to chance in a 3-way forced choice task.  
Thus, the subject's discrimination performance is significant if: 

\begin{equation}
\label{eq:analytical1}
\alpha/2 > 1 - cdf_{binomial}(T,k,\frac{1}{3}) = \sum_{i=0}^{k} \binom{T}{i}(\frac{1}{3})^{i}(\frac{2}{3})^{T-i}
\end{equation}

For $\alpha=0.05$, $T=20$ (as used in \cite{bushdid_humans_2014}), this inequality is satisfied for $k>=11$.  
For each subject, $k$ might be any value between 0 and 20 depending on olfactory acuity.  
If $k>=11$ for more than $50\%$ of subjects, 
then the value of $D$ characterizing that mixture pair is necessarily $>d$.  
If $k>=11$ for fewer than $50\%$ of subjects, then $D<d$.  
If $k>=11$ for exactly $50\%$ of subjects, then $D=d$.  
The actual estimate for $d$ is obtained by regression in the spirit of Figure \ref{fig:reconstruction}.  

What kind of subject can discriminate successfully 11 times out of 20?  
Consider a mixture class $X_{N,D}$ (characterized by $N$ and $D$), 
and a subject performance of $f_{N,D}$, 
corresponding to the proportion of mixtures correctly discriminated from a sample of size $T$.  
Note that $f_{N,D}$ is simply the abscissa of Figure 1 from \cite{bushdid_humans_2014}.   
A subject with $f_{N,D}=0.55$ would get $k=T*f_{N,D}=11$ out of $T=20$ correct on average.  
So we can rewrite the inequality above as an equation:  

\begin{equation}
\label{eq:analytical2}
1 - \alpha/2 = \sum_{i=0}^{f_{N,D}*T} \binom{T}{i}(\frac{1}{3})^{i}(\frac{2}{3})^{T-i}
\end{equation}

If the above equation is satisfied, then the subject will be considered to be on the boundary 
between significantly discriminating and not significantly discriminating mixture pairs in the class.  
If half of subjects perform better than $f_{N,D}$, and half less, 
then half of subjects will be considered to significantly discriminate mixture pairs in the class (and half not), 
and so $d$ will be set equal to $D$.  
This is simply the definition of $d$.

The value $f_{N,D}$ for which that equation is satisfied depends upon $\alpha$ and $T$.  
$f_{N,D}$ is related to $N$ and $D$ through the data, 
and so the value of $D$ for which the equation is satisfied (i.e. $D=d$)
depends upon $\alpha$, $T$, and the data.  
However, it is inappropriate for the discriminability limen to depend on $\alpha$ and $T$ in this way.  
As we showed above, this has serious consequences for the estimate of $d$, and therefore also for the estimate of $z$.  
It is what makes $z$ inconsistent.  

\begin{figure}[!hbt]
    \centering
    \includegraphics[width=0.475\textwidth]{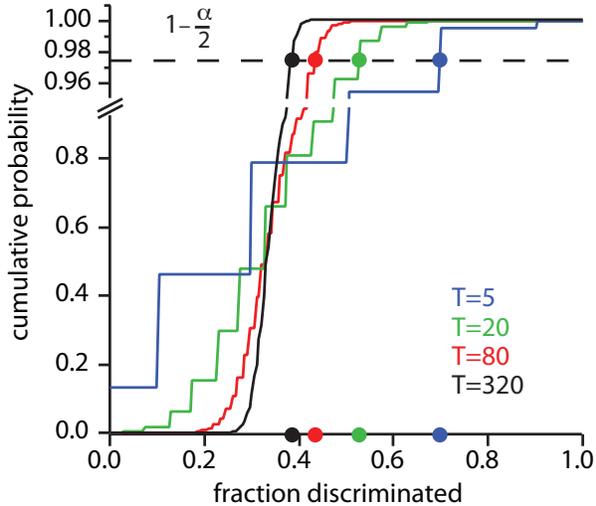}
    \caption{
Fraction discriminated at which statistical significance is reached. For each possible value of the number of tests $T$ conducted per mixture class, there is a cumulative distribution of the fraction $f$ of those tests that will be correctly discriminated, under the null hypothesis of chance ($\frac{1}{3}$) responding.  The choice of significance threshold $\alpha$ determines the fraction correct required to reject the null hypothesis, and thus count as `significantly discriminating' in the framework.  For a given value of $\alpha$ (0.05 shown here, and used in \cite{bushdid_humans_2014}), the fraction correctly discriminated required to reach this threshold varies greatly with $T$. Rejecting the null hypothesis can thus be very easy or very hard depending on $T$ (or the number of subjects $S$, not shown), or on $\alpha$.}
    \label{fig:alpha_crossing}
\end{figure}

Figure \ref{fig:alpha_crossing} shows the relationship between the critical $f_{N,D}$, $T$, and $\alpha$.  
Note that this relationship is independent of the data.  
The data only determine how $f_{N,D}$ depends upon $D$ and consequently determines $z$.  
In summary, a smaller (larger) value of $\alpha$ or $T$ requires a much higher (lower) value of $f_{N,D}$ to satisfy the equation.  This higher (lower) value of $f_{N,D}$ might only be found at a much larger (smaller) value of $D$, implying a much larger (smaller) value of $d$ and therefore a much smaller (larger) value of $z$.  

With a sufficient number of subjects (or tests), 
even barely above chance performance can produce estimates of $z$ equal to the largest possible number of stimuli (Figures \ref{fig:colormaps} and \ref{fig:alpha_dependency}).  
In fact, this is guaranteed by equation \ref{eq:analytical2}.  
The critical values of $f_{N,D}$ required for statistical significance will asymptotically approach $\frac{1}{3}$ (chance)
as $T$ approaches infinity.  
The same principle applies to a consideration of changes to the number of subjects $S$, 
instead of the number of tests.  
This illustrates the core of the problem.  
Discriminating significantly above chance can be a very high bar or a very low bar 
depending on the parameters of the experiments and the analysis, including $S$, $T$, and $\alpha$.  

\section{Acknowledgements}
We thank Krishnan Padmanabhan and Shreejoy Tripathy for helpful comments on the manuscript.

{\footnotesize \bibliographystyle{acm}
\bibliography{references}}

\end{document}